\documentclass[prl,twocolumn,tightenlines,showpacs,nofootinbib]{revtex4}
\usepackage{bm,dcolumn,amsmath,graphicx}
\usepackage
{hyperref}

\begin{document}
\title{Electric dipole moment enhancement factor of thallium}
\author{S. G. Porsev$^{1,2}$}
\author{M. S. Safronova$^1$}
\author{M. G. Kozlov$^2$}
\affiliation{$^1$ Department of Physics and Astronomy, University of Delaware,
                  Newark, Delaware 19716, USA}
\affiliation{$^2$ Petersburg Nuclear Physics Institute, Gatchina,
                  Leningrad District, 188300, Russia}
\date{\today}
\begin{abstract}

The goal of this work is to resolve the present controversy in the value of the EDM enhancement factor in Tl. We have carried out several
calculations by different high-precision methods, studied previously omitted corrections, as well as tested our methodology on other parity
conserving quantities. We find the EDM enhancement factor of Tl to be equal to $-$573(20). This value is 20\% larger than the recently published
result of Nataraj {\it et al.} [Phys.~Rev.~Lett. {\bf 106}, 200403 (2011)], but agrees very well with several earlier results.

\end{abstract}

\maketitle
 A number of extensions of the standard model of particle physics predict electric dipole moments (EDM) of particles that may be observable with the present
state-of-the art experiments \cite{SM1} making EDM  studies a remarkable tool in search for new physics. The EDMs arise from the  violations of both
parity and time-reversal invariances. The present constraints on the EDMs are already within bounds predicted by some theories \cite{SM1}. If the
EDMs are not observed in the next generation of the experiments, some of the low-energy supersymmetry and other theories will be ruled out. The
standard model predicts tiny electron EDM, $d_e <10^{-40}~ e$~cm since it can not originate even from three-loop diagrams \cite{KhrLam97}.

 The
electron EDM is enhanced in certain atomic and molecular systems, and two of the most stringent limits on the electron EDM $d_e$ were obtained from
the experiments with $^{205}$Tl: $d_e<1.6\times 10^{-27}e$~cm \cite{RCS02}, and with YbF molecule: $d_e<1.05\times 10^{-27}e$~cm \cite{HKSS11}. These
limits significantly constrain supersymmetric and other extensions of the standard model \cite{SM1}.

 Both results crucially depend on the calculated
values of the effective electric field on the valence electron. In the case of Tl this effective field is proportional to the applied field $E_0$,
$E_\mathrm{eff}=K E_0$, and $d_\textrm{at}(^{205}\textrm{Tl})=K d_e$. The quantity $K$ is referred to as the EDM enhancement factor.

 Until recently,
there was a consensus that the value of $K$ is close to $-580$ \cite{LiuKel92,DzuFla09}, but the latest calculation \cite{NatSahDas11} gave the value
$-466(10)$, or more than 20\% smaller. All three of these calculations  used high-accuracy methods that include some parts of the correlation
corrections to all orders.   Liu and Kelly ~\cite{LiuKel92} used relativistic coupled-cluster approach, but had to make various restrictions in their
calculations to make it manageable with the computer power available in 1992. Dzuba and Flambaum \cite{DzuFla09} used a combination of configuration
interaction method (CI) with many-body perturbation theory (MBPT) starting from the [Xe]$4f^{14}5d^{10}$ Dirac-Fock (DF) potential and considering
thallium as a system with three valence electrons. This potential is referred to as $V^{N-3}$, where $N$ is the total number of electrons. Nataraj
{\it et al.} \cite{NatSahDas11} used relativistic coupled-cluster (RCC) method with single, double and perturbative triple excitations of the
Dirac-Fock wave functions starting from the [Xe]$4f^{14}5d^{10}6s^2$ potential. In this potential, referred to as $V^{N-1}$, $6s^2$ shell is included
in the core and thallium is considered to be  a monovalent system, such as an alkali-metal atom.  Both relativistic coupled-cluster method, in its
various implementations, and CI+MBPT method have been used for a number of years in many other applications, including study of parity violation,
calculation of other (P, T-odd) effects, search for variation of fundamental constants, and many others.
 We note that calculations of the effective field for such a heavy molecule as YbF is more difficult than for atomic Tl, and discrepancy
in the theory for atoms may compromise molecular limit \cite{HKSS11} as well.

Because of the importance of this issue, we return to the problem of Tl EDM in this letter. We have carried out several calculations by different
high-precision methods in different potentials. Below, we briefly summarize the calculations that we have performed and our main findings before
providing more details of the methods used in this work.

\textbf{1. CI+MBPT calculation in $V^{N-3}$ potential.} Firstly, we have repeated the calculation of Dzuba and Flambaum \cite{DzuFla09} and ensured
that we agree with their value at the same level of approximation.   Then, we have calculated a number of corrections that were omitted in
\cite{DzuFla09}, including structure radiation, core-Brueckner, two-particle, selected three-particle, and normalization corrections. We found that
some of these corrections are large, $5-7$\% percent, but partially canceling, causing the lower accuracy of $V^{N-3}$ results than was previously
expected.

\textbf{2. CI+all-order calculation in $V^{N-3}$ potential.} Recently, we have developed the relativistic  method \cite{SafKozJoh09}  combining CI
with relativistic coupled-cluster (CC) approach. This method, first suggested in \cite{Koz04},  (here referred to  as the CI+all-order method) was
successfully applied to the calculation of divalent atom properties in Refs.~\cite{SafKozJoh09,PRLBBR}. This calculation allowed us to evaluate the
effect of higher-order core-core and core-valence corrections to the EDM. We found that  the effect of these correction to the enhancement factor is
small, 0.7\%.

\textbf{3. CI+MBPT calculation in $V^{N-1}$ potential.} We have repeated the entire CI+MBPT calculation described above, including all corrections,
in $V^{N-1}$ potential. We note that  the CI+MBPT approach still allows us to fully treat all three valence electrons using CI. Therefore, we can
accurately treat the contributions of the $6s6p^2$ states to $K$ on the same footing as $6s^2ns$ terms unlike the approach of \cite{NatSahDas11}. Our
final result is based on the V$^{N-1}$ calculation with the higher-order corrections estimated from the V$^{N-3}$ CI+all-order calculation.

\textbf{4. RCC calculation of the $6s^2ns$ contributions to EDM.} In an attempt to compare with RCC calculations of \cite{NatSahDas11}, we carried
out full relativistic coupled-cluster calculations including single, double, and perturbative triple (RCCSDpT) excitations of the DF functions in
$V^{N-1}$ potential. We have also used this calculation to evaluate the effect of the Breit interaction to EDM and found it to be negligible
(0.36\%). We have verified that our RCCSDpT value for the $6s^27s$ and $6s^28s$ contributions to the EDM are in excellent (2\%) agreement with our
final CI+MBPT contribution confirming the accuracy of our calculations. We have conducted RCC calculations with a truncated basis that we have
constructed using parameters of \cite{NatSahDas11}. The truncation of the basis to the size of Ref.~\cite{NatSahDas11} produced very large reduction
(by 18\%) in the size of the $6p_{1/2}-7s$ EDM matrix element.

We discuss these four calculations in more details below. We start with solving  DF equations $ \hat H_0\, \psi_c = \varepsilon_c \,\psi_c, $ where
$H_0$ is the relativistic DF Hamiltonian \cite{DzuFlaKoz96b,SafKozJoh09} and $\psi_c$ and $\varepsilon_c$ are single-electron wave functions and
energies. The self-consistent calculations were performed for the [$1s,...,5d^{10}$] closed core  and the $6s-8s$, $6p$, $7p$, and $6d$ orbitals were
obtained in $V^{N-3}$ approximation.  We have constructed the basis set \cite{Bog91,KozPorFla96} consisting of 166 orbitals ($22s$, $22p$, $21d$,
$20f$, $13g$, and $11h$). In order to estimate the accuracy of this basis set, we repeated some of the calculations with significantly larger basis
set consisting of 273 orbitals ($35s$, $34p$, $28d$, $27f$, $21g$, and $20h$) and found that the differences were small and well below our estimated
accuracy.  The CI space was significantly increased in comparison to \cite{DzuFla09}, and included ($22s$, $22p$, $17d$, and $16f$) orbitals; higher
$n$ orbitals were allowed fewer number of excitations. Such CI space is effectively complete. All MBPT and all-order calculations were carried out
with inclusion of all orbitals.

The multiparticle relativistic equation for three valence electrons is solved within the CI framework \cite{KT87} to find the wave functions and the
low-lying energy levels: $ H_{\rm eff}(E_n) \Phi_n = E_n \Phi_n,$ with  the effective Hamiltonian defined as $ H_{\rm eff}(E) = H_{\rm FC} +
\Sigma(E).$  $H_{\rm FC}$ is the Hamiltonian in the frozen-core approximation and the energy-dependent operator $\Sigma(E)$ takes into account
virtual core excitations. The  $\Sigma(E)$ part of the effective Hamiltonian is constructed using the second-order perturbation theory in the CI+MBPT
approach \cite{DzuFlaKoz96b} and linearized single-double coupled-cluster method (LCCSD) in the CI+all-order approach \cite{SafKozJoh09}. Since the
valence-valence correlations are very large, the CI method provides better description of these correlations than the perturbative approaches such as
RCC due to possible large contributions of higher-order (or higher-excitation) corrections.  The LCCSD method used here is known to describe the
core-core and core-valence correlation very well as demonstrated by its great success in prediction of alkali-metal atom properties \cite{reviewall}.
Therefore, combination of the CI and all-order LCCSD methods allows to account for all dominant correlations to all orders.

The absolute values of the three-electron binding energy and the energy levels of the low-lying excited states in respect to the ground state
obtained in the pure CI, the CI+MBPT, and the CI+all-order approximations are given in Table I of the supplementary material~\cite{suppl}.
 We find that the CI+all-order improves the accuracy of energies and reduces the error in the ground state three-electron binding energy to 0.2\% level.

The atomic EDM ${\bf d}_{\rm at}$ of the ground state of Tl is defined as
\begin{equation}
{\bf d}_{\rm at} = 2 \sum_n \frac{\langle 0|{\bf D}|n \rangle \langle n|H_d|0 \rangle}{E_0 - E_n}, \label{d_at}
\end{equation}
where ${\bf D}$ is the electric dipole moment operator. The operator $H_d$ is given by \cite{KhrLam97}:
\begin{equation}
H_d = 2 d_e
\left(
\begin{array}{rr}
 0 & 0 \\
 0 & {\bm \sigma} {\bf r}
\end{array} \right)
\frac{Z(r)}{r^3},
\label{Hd}
\end{equation}
where $d_e$ is the EDM of the electron, $Z(r)$ is the charge of the nucleus and core electrons within the sphere of radius $r$, and ${\bm \sigma}$
are Pauli matrices. In the CI+MBPT approach, we construct effective valence operators for all observables of interest
\cite{DzuKozPor98,PorRakKoz99P}. In this work, we need effective operators for the electric dipole operator $D_{\rm eff}$, magnetic-dipole hyperfine
(hfs) interaction $(H_\mathrm{hfs})_{\rm eff}$, and the operator $({H_d})_{\rm eff}$. These operators account for the core-valence correlations in
analogy with the effective Hamiltonian. We do not perform explicit summation over three-particle states in our approach, but use
Dalgarno-Lewis-Sternheimer method that involves solution of the inhomogeneous equation with the corresponding effective operators
~\cite{DzuKozPor98,PorRakKoz99P}.   We include additional corrections beyond random-phase approximation (RPA) in the construction of all effective
operators in comparison with ~\cite{DzuFla09}. These contributions include  the core-Brueckner ($\sigma$), two-particle (2P) corrections, structural
radiation (SR), and normalization (norm) corrections. Finally, we calculated selected three-particle (3P) corrections to the effective
Hamiltonian~\cite{DzuFlaKoz96b}.
\begin{table}
\caption{The ground state three-electron binding energy $|E_{\rm v}|$ (in a.u.) and the energy levels of the low-lying excited states in respect to
the ground state (in cm$^{-1}$) for $V^{\rm N-1}$ approximation. Results of the calculations and the differences from the experimental values
\cite{RadSmi85,Moo71} (in \%) are presented for CI and CI+MBPT approximations.} \label{Tab:Ev1}
\begin{ruledtabular}
\begin{tabular}{lccccc}
           & \multicolumn{2}{c}{CI}
                        & \multicolumn{2}{c}{CI+MBPT}
                                       & Expt.\footnotemark[1] \\
\hline
$E_{\rm v}$ & 1.9809&$  4\%$& 2.0682&$  0.2\%$& 2.0722\\
\hline
$6p_{3/2}$  &  7016 &$ 10\%$& 7854  &$ -0.8\%$&  7793 \\
$7s_{1/2}$  & 24649 &$  7\%$& 26328 &$  0.6\%$& 26478 \\
$7p_{1/2}$  & 31876 &$  7\%$& 33954 &$  0.6\%$& 34160 \\
$7p_{3/2}$  & 32834 &$  7\%$& 34974 &$  0.5\%$& 35161 \\
$6d_{3/2}$  & 33762 &$  7\%$& 36106 &$  0.0\%$& 36118 \\
$6d_{5/2}$  & 33828 &$  7\%$& 36180 &$  0.1\%$& 36200 \\
$8s_{1/2}$  & 36549 &$  6\%$& 38693 &$  0.1\%$& 38746
\end{tabular}
\end{ruledtabular}
\end{table}
We find that an accurate calculation of different observables in $V^{\rm N-3}$ potential is more complicated due to the poor convergence of the MBPT.
We present the contributions to hfs constants $A$ for the 8 low-lying states in the Table II of the supplementary material~\cite{suppl}. We find that
many corrections beyond CI+MBPT and RPA are large and partially canceling. As a result, an agreement between final theoretical values and the
experimental results in certain cases is not very good. In particular, the discrepancy between theoretical and experimental values of $A(7s)$ is at
the level of 8\%.  The normalization corrections are unusually large ($\sim$ 6\%). We calculated the normalization correction by  approximately
expressing it in terms of the derivatives of the MBPT corrections in respect to the energy \cite{DzuFlaKoz96b}. It appears that different method for
treatment of this correction needs to be developed in the case of $V^{\rm N-3}$ potential.

\begin{table}
\caption{The magnetic-dipole hfs constants (in MHz) and the absolute values of the reduced matrix elements of the electric-dipole operator $|\langle
\gamma ||D|| \gamma' \rangle|$ (in a.u.)} \label{Tab:hfs_E1}
\begin{ruledtabular}
\begin{tabular}{lccr}
       &            & Theory & \multicolumn{1} {c} {Expt.} \\
\hline
$A$ (MHz) & $6p_{1/2}$ &  22041 & 21310.8~\cite{hfs6p}\\
          & $7s_{1/2}$ &  12395 & 12297(2)~\cite{hfs}\\
          & $8s_{1/2}$ &  3900  & 3871(1)~\cite{hfs} \\
E1 (a.u.) & $|\langle 7s ||D|| 6p_{1/2} \rangle|$ & 1.781 & 1.81(2)~\cite{e1} \\
          & $|\langle 8s ||D|| 6p_{1/2} \rangle|$ & 0.521 &  \\
          & $|\langle 7s ||D|| 6p_{3/2} \rangle|$ & 3.393 & 3.28(4)~\cite{e1} \\
          & $|\langle 8s ||D|| 6p_{3/2} \rangle|$ & 0.764 &
\end{tabular}
\end{ruledtabular}
\end{table}
We find the same problem when calculating these correction to the  EDM enhancement factor  in the $V^{\rm N-3}$ approximation.
 The CI value is $-584$ and the CI+MBPT, CI+all-order, and RPA corrections contribute only 3, 4, and 3, respectively. Usually these are
the most important corrections to the valence CI. At the CI+MBPT+RPA level, our result  is $-578$ and is in a good agreement with the value
$-582(20)$ obtained by Dzuba in Flambaum using the same CI+MBPT+RPA approximation in the $V^{\rm N-3}$ potential. Small difference may be due to a
different basis set and larger CI space (including $l=3$ partial wave) in our calculations. The corrections $\sigma$, SR, 2P, 3P, and norm are $25$,
$-1$, $-22$, $-2$, and $36$. The two-particle  and normalization corrections are large, $+4\%$ and $-6\%$, correspondingly leading to the value
$K=-538(46)$. We estimated the uncertainty in $K$ based on the difference of the relevant hyperfine constants with experiment (8\% for $A(7s)$) and
the total contribution of all corrections beyond CI (8.6\%).

In summary, we find that the corrections beyond CI+all-order+RPA are large; even though they partially cancel each other, their total
 contribution is significant, almost 7\% in $V^{\rm N-3}$ potential. At the same time, the all-order CC corrections due to
higher-order core-valence correlations are very small, 0.7\%. We conclude that the size of different corrections to the EDM in the V$^{\rm N-3}$
approximation is not typical and missing higher-order contributions to the effective operators can be important. Because of that, we repeat
calculations in the V$^{\rm N-1}$ approximation. We already used this approximation in the calculation of the parity-nonconserving amplitude for the
$6p_{1/2}-6p_{3/2}$ transition in Tl with 3\% accuracy \cite{KozPorJoh01}. Comparison of the $V^{\rm N-1}$ and $V^{\rm N-3}$ potentials for Tl
calculations has been recently discussed in Ref.~\cite{new}.
\begin{table*}[tb]
\caption{The breakdown of different contributions to our final value of the EDM,  V$^{\rm N-1}$ potential. First column gives the CI value and the
following columns give various corrections described in the text.} \label{Tab:VN1EDM}
\begin{ruledtabular}
\begin{tabular}{cccccccccccc}
  CI    & CI+MBPT & RPA   & Sbt  & 2P    & $\sigma$ & SR  & Norm  & Final & Ref.~\cite{LiuKel92} & Ref.~\cite{DzuFla09} & Ref.~\cite{NatSahDas11}  \\
\hline
 -593.6 &  8.7    & -13.0 & 16.5 & -18.8 &  22.5    &  0.0  & 5.2 & -573(20)  &-585(30-60) &-582(20) &-466(10)
\end{tabular}
\end{ruledtabular}
\end{table*}

CI+MBPT calculation in the V$^{\rm N-1}$ potential follows the same procedure as the one in the V$^{\rm N-3}$ approximation, but the self-consistency
DF procedure is carried out for the [$1s,...,5d^{10}, ~6s^2$] core. We note that we use the Brillouin-Wigner variant of the MBPT in both cases. In
this formalism, the effective Hamiltonian for the valence electrons is energy-dependent. It was shown in our work~\cite{KozPor99O} that the accuracy
of the theory can be improved by calculating the Hamiltonian at the optimal valence energy for Tl, which  was found to be $-1.8$~a.u.. In
Table~\ref{Tab:Ev1}, we present the absolute values of the valence energy of the ground state and the energy levels of the low-lying excited states
counted from the ground state obtained in the pure CI and in the CI+MBPT approximations. We note that the CI+all-order formalism is presently limited
to the $V^{\rm N-3}$ potential. In $V^{\rm N-1}$, so-called subtraction diagrams have to be included consistently at the all-order level which so far
has not been implemented. Since the all-order core-valence corrections contributed only 0.7\% in the $V^{\rm N-3}$ approximation, these are small at
the present calculation as well. A comparison of the results presented in Tables~\ref{Tab:Ev1}  shows that the energy levels found in V$^{\rm N-1}$
approximation turn out to be closer to the experimental results than in V$^{\rm N-3}$ approximation, which is already observed at the stage of pure
CI approximation. As a result, the MBPT corrections that give the main contribution to the uncertainty budget are smaller leading to better agreement
between the theoretical and experimental energy levels. Our values for the  magnetic-dipole hfs constants and E1 transition amplitudes between
low-lying levels in V$^{\rm N-1}$ approximation are compared with experimental results  \cite{e1,hfs6p,hfs} in Table~\ref{Tab:hfs_E1}. The
calculation of these properties was discussed in detail in Ref.~\cite{KozPorJoh01}. The corrections that are taken into account are similar to those
for the V$^{\rm N-3}$ approximation. The only essential difference is an appearance of the subtraction diagrams (Sbt) in the former case.  The
differences between the theoretical and experimental results do not exceed 3\% for all relevant properties.

In Table~\ref{Tab:VN1EDM}, we present the breakdown of different contributions to the atomic EDM. The difference between the CI value, $-594$, and
the final value, $-573$, is only 3.7\%. It demonstrates that the interaction between valence electrons is much more important and should be treated
as accurately as possible.  In Table~\ref{Tab:EDMval}, we list  the most significant contributions of certain valence states that were calculated
using
 Eq.~(\ref{d_at}) and results of our  $V^{\rm N-1}$ CI+MBPT calculation. These contributions give 85\%  of the total value.
\begin{table}
\caption{The contributions  to the EDM  in our final $V^{\rm N-1}$ CI+MBPT calculation.  Columns $D$ and $H_d$ give reduced matrix elements of the
electric-dipole and EDM operators. The results of our RCCSDpT calculation for the corresponding $6s^2 ns$ contributions are given in the rows labeled
$K$(CC) for comparison.} \label{Tab:EDMval}
\begin{ruledtabular}
\begin{tabular}{lcccrrr}
         State        &$\Delta E_\textrm{th}$ &$\Delta E_\textrm{expt}$&
                                        $D$
                                                 & $H_d$
                                                         &  \multicolumn{1}{c}{$K$} &$K$(CC)\\
                                                         \hline
$6s^2 7s~^2\!S_{1/2}$    & 26328 &  26478  & -1.798 & 17.7 & -216.6 & -212.2  \\
$6s^2 8s~^2\!S_{1/2}$    & 38693 &  38746  & -0.526 &  9.2 &  -22.4 & -22.8\\
$6s 6p^2~^4\!P_{1/2}$    & 46281 &  45220  & -0.427 & 45.1 &  -74.6 &\\
$6s 6p^2~^2\!P_{1/2}$    & 69218 &  67150  & -2.472 & 28.0 & -179.2 &\\
$6s 6p^2~^4\!D_{1/2}?$ & 79830 &         & -0.142 & -4.2 &    1.3 &\\
Other                   &        &       &         &     & -81.1 &\\
 Total                        &       &         &        &  & -572.5&
\end{tabular}
\end{ruledtabular}
\end{table}
Table~\ref{Tab:EDMval} illustrates that it is very important to accurately account for  the contributions of  the $6s 6p^2$ configurations. Their
contribution to the EDM is $\sim$ 45\%.  In the RCC method of \cite{NatSahDas11}, these contribution are accounted for as the excitations of the core
electrons which is unlikely to provide required accuracy.

We have also carried out completely different set of calculations using relativistic coupled-cluster method with single, double, and perturbative
triple (RCCSDpT) excitations in $V^{\rm N-1}$ potential ~\cite{reviewall,NL} to evaluate the dominant contributions to the EDM from the $6s^27s$ and
$6s^28s$ states by a different approach. All nonlinear terms were included at the SD level. This method is theoretically very close to that of
\cite{NatSahDas11}. While there are differences in the treatment of triple excitations between the two approaches, we find that contributions of the
triple excitations to the EDM is small (less than 2\%). We have also used this calculation to evaluate the effect of the Breit interaction to EDM and
found it to be negligible (0.36 \% for the $6s^2 7s$ contribution). Our final values for the $6s^2 7s$ and $6s^2 8s$ contributions are given in the
last column of Table~\ref{Tab:EDMval}. RCCSDpT values are in excellent (2\%) agrement with our final CI+MBPT values. The agreement of the results
obtained by two completely different approaches confirms the accuracy of our calculations. However, the value of this contribution inferred from
Fig.~2 of \cite{NatSahDas11} is 10\% lower, about $-188$. We find that this difference may be due to significant truncation of the basis set used in
RCC calculation of \cite{NatSahDas11}. The main part of our RCCSDpT calculation was carried out with very large numerically complete basis set (650
orbitals with $l<7$). Fig.~2 in \cite{NatSahDas11} shows that their $9s$ orbital already belongs to continuum, which is due to use of only $ns$
orbitals up to $14s$ in the RCC calculation. We have conducted a basis set test truncating our 166 orbital basis to $n=14$ for all partial waves and
using it in the RCC calculations. We find that the value of the $6p_{1/2}-7s$ EDM matrix element was reduced by 18\% due to basis set truncation.

To conclude, we calculated the EDM enhancement factor to be equal to $-$573(20). The uncertainty is, somewhat conservatively,  assigned based on the
accuracy of the relevant hyperfine constants (3\%)  and total size of \textit{all} corrections beyond the CI, which is 3.7\%. This value differs by
20\% from the recently published result of Nataraj {\it et al.} \cite{NatSahDas11}, but agrees well with the results obtained by~\citet{DzuFla09}
and~\citet{LiuKel92}.

The authors thank A. Derevianko, V. Dzuba, and V. Flambaum for helpful discussions. This work was supported in part by US NSF Grants No.\ PHY-1068699
and No.\ PHY-0758088. The work of MGK was supported in part by RFBR grant \#11-02-00943.


\end{document}